\documentclass{article}

\usepackage{arxiv}
\usepackage{authblk}
\usepackage{amsmath}

\title{On Probabilistic Byzantine Fault Tolerance}
\author{Quan Nguyen, Andre Cronje}
\affil{FANTOM}

\begin{document}
\maketitle

\begin{abstract}
Byzantine fault tolerance (BFT) has been extensively studied in distributed trustless systems to guarantee system's functioning when up to 1/3 Byzantine processes exist. 
Despite a plethora of previous work in BFT systems, they are mainly concerned about common knowledge deducible from the states of all participant processes. In BFT systems, it is crucial to know about which knowledge a process knows about the states of other processes and the global state of the system. However, there is a lack of studies about common knowledge of Byzantine faults, such as, whether existence of a Byzantine node is known by all honest nodes. In a dynamic setting, processes may fail or get compromised unexpectedly and unpredictably. It is critical to reason about which processes know about the faulty processes of the network.

In this paper, we are interested in studying BFT systems in which Byzantine processes may misbehave randomly, possibly at some random periods of time. The problem of \emph{probabilistic Byzantine} (PB) processes studied in this paper is more general than the problems previously studied in existing work.
We propose an approach that allows us to formulate and reason about the concurrent knowledge of the PB processes by all processes. We present our study of the proposed approach in both synchronous and asynchronous systems. 
\end{abstract}


\section{Introduction}\label{ch:intro}

There have been an upsurge of interests in 
cryptocurrencies and distributed ledger technologies since the success of Bitcoin. At its core, a blockchain system relies on a consensus protocol that ensures all nodes in the network agree on a single chain of transaction history. The functioning is guaranteed, regardless of any possible adverse influence of malfunctioning and malicious nodes. The underlying blockchain technologies have seen a vast amount of interests for business and innovation, with applications ranging from logistics, healthcare, and smart cities.

Fault-tolerant consensus has been extensively studied in distributed systems~\cite{attiya1984asyn,Castro99,abraham2017efficient,mostefaoui2015sig}. \emph{Byzantine} fault tolerance (BFT)~\cite{Lamport82} is tolerant up to a third of the participant machines in failure. In BFT systems, reaching agreement is crucial to guarantee consistency of the continuous replication of distributed state machine across the network. 
Byzantine nodes refer to participant processes (or machines) which may be at fault, get disconnected or some of them can be adversary. 
A Byzantine failure is often caused by a malfunctioning or malicious process~\cite{Lamport82}.  If multiple Byzantine components co-exist, they may collude to cause more damage to the network. Byzantine fault is considered the most severe and challeging to deal with, and crash failure is often considered a benign case.

In spite of different states in distributed processes, a BFT consensus algorithm guarantees all (honest) processes agree on common states and common data values. Services built on top of these BFT algorithms can guarantee that all honest nodes will perform same sequence of actions regardless of faulty components and unreliable communication links. This consensus guarantee is crucial in distributed BFT systems. 
Consensus algorithms, which guarantee transaction integrity over the distributed network, are equivalent to the proof of BFT~\cite{randomized03, paxos01}. 
Practical BFT (pBFT)~\cite{Castro99} can reach a consensus for a block once the block is shared with other participants and the share information is further shared with others \cite{zyzzyva07, honey16}.

Most of the previous work in BFT systems have mainly focused on \emph{common knowledge} or \emph{common state} of the states of all participant processes in the network~\cite{cck92,dwork1986knowledge}. This is fundamentally important in BFT systems, to know about which knowledge a process has learned about the other processes' states, as well as the global state of the system. However, there is a gap in justifying about common knowledge of Byzantine faults in a BFT system, such as, whether existence of a Byzantine node is known by all honest nodes. Giv en the dynamic nature of a realistic network, processes may fail or get compromised unexpectedly and unpredictably. Thus, it is critical to reason about which processes know about the faulty processes of the network.

To motivate our study, we start with the following problem, which describes a simple version of Byzantine fault detection problem.

\subsection{The cheater problem}\label{sec:cheater}

In a remote island, there are $n$ villagers living together.
Each day, all villagers come out to collect fruits around the island and carry them back in a box back to the village. Each villager individually counts the number of fruits in the box and records their count on the box. They can then check if there is a count that majority of them agree on. Then s/he can find out whether there is a cheater(s). 
If a cheater is found, each villager then gossips the new finding to all villagers on the same day.

The objective of the problem is to know if the villagers can discover and agree on the existence of every cheater in the village. Previous work has mainly focused on ensuring that all honest villagers do know the correct number of fruits (i.e., a common knowledge of states of all processes). There is a lack of understanding if they agree whether all cheaters are known by the honest villagers. It also remains unknown whether the common knowledge about cheaters will be possible if cheaters have different probability of cheating.

\subsection{Our approach}

In order to study about the Cheater problem, we investigate a model of BFT system in which individual cheaters may have different probabilities of cheating. There are two types of nodes in a distributed system: Zenta (honest) nodes and Byzantine (cheating) nodes.
In our new approach, we assume that every Byzantine $i$ (cheater) has a probability $\epsilon_i$ of fault (cheating), whereas every Zenta nodes have a zero probability. 
Based these probabilities, we then study whether the network will ultimately discover all Byzantine processes (cheaters).

In particular, this paper introduces a semi-formal model of the cheater problem.
We use some terminologies of network, process and common knowledge, as in previous work~\cite{cck92}.
A \emph{process} is a participant machine or node of the network. There are two types of processes: (1) Zenta process is a honest process (denoted by Z); (2) Byzantine process is a malfunctioning or malicious process (denoted by B). 
Unlike previous work, we consider a general case of Byzantine processes, which can misbehave randomly and unpredictably. These are so-called \emph{probabilistic Byzantine} (PB) processes.
In its most general form, our model considers each process $i$ (whether Z or B) has a probability of cheating $\epsilon_i$ $\in$ [0,1]. A Z process has $\epsilon_i$ = 0, i.e., its probability of cheating is zero. In contrast, a B or PB process has a positive probability of cheating, e.g., $0< \epsilon_i \leq 1$.

We then present our study of the proposed model
in both synchronous and asynchronous BFT systems. 
We show how the model can help reason about common knowledge of the probabilistic Byzantine processes by all process in the network. Interestingly, our study of the model has shown that a PB with a higher probability of cheating is more likely to be detected than other PBs with a lower probability. 

The rest of the paper is organized as follows. Section~\ref{se:related} gives related work.
Section~\ref{se:ourapproach} decribes our approach to analyse the common knowledge of all processes with respect to the cheating values $\epsilon_i$. 
Section~\ref{se:asyn} gives our study of asynchronous BFT systems using our approach.
Section~\ref{se:discuss} covers several discussions followed from our studies. 
Conclusion is given in Section~\ref{se:con}.

\section{Related work} \label{se:related}

This section gives related work on fault-tolerant consensus and Byzantine fault-tolerance in distributed systems.

\subsection{Byzantine fault tolerance in trustless systems}
A distributed system is often comprised of physically distinct entities, which are also named nodes, processors, agents or sensors. 
They are geographically separated and each of them has only a partial knowledge of the system. The term \emph{process} is commonly used to denote any computing entity. The system is functional if every processes of a system exchange information, and can reach agreement with each other on certain data values to achieve a common goal.
The system can face unpredictable faults and adversarial influence due to faulty processes and unreliable communication channels.

Byzantine fault tolerance has been studied extensively. Examples include Practical BFT (pBFT)~\cite{Castro99}, Paxos~\cite{paxos01},  Zyzzyva~\cite{zyzzyva07}, Q/U~\cite{abd2005fault}, HQ~\cite{cowling2006hq}, just to name a few. Two types of failures are crash failure and Byzantine failure. In Byzantine failure, the process may act arbitrarily, send contradictory messages to peers or simply remain silent. Whilst a crashed process stops functioning completely and does not resume. 
BFT systems can be synchronous~\cite{abraham2017efficient}, partially asynchronous, or asynchronous~\cite{mostefaoui2015sig}.

For asynchronous distributed systems~\cite{attiya1984asyn,fantom18,stakedag}, it is challenging due to the very nature of distributed computing in that each process knows only a partial knowledge of the system, and none can capture instantaneously the global state of the system. This is because of geographical distance between the processes and presence of uncertainty in asynchrony and failures. Theoretically, a crashed process is impossibly distinguished from a very slow processs in an asynchronous BFT system.

BFT protocols have been used in a wide range of applications including replicated file system, backup, and block stores. Many of them guarantee~\emph{safety} and ~\emph{liveness}, even though arbitrarily Byzantine replicas may exist. The safety property~\emph{linearizability} ensures a sequential order of execution of the requests as seen by clients, whereas the liveness property ensures that all valid  requests from clients are eventually executed.

\subsection{Knowledge and Common knowledge}
There are several work that give a fundamental understanding about knowledge and common knowledge in distributed systems. CCK paper~\cite{cck92} defines a formal model of concurrent common knowledge, which is used to study in asynchronous systems. Common knowledge is also studied in Byzantine environment \cite{dwork1986knowledge}.

Byzantine consensus considers the problem of reaching agreement among a system of $n$ processes $p_1$, $\dots$, $p_n$ ($n > 2$). The processes communicate by sending messages to each other. Each process $i$ has an initial value $v_i$, and it has to decide on a value $v$ at each step. 
A BFT system may contain up to $f$ Byzantine processes which may deviate from the protocol in an arbitrary manner. Regardless of the determinism of the protocol, several factors can cause non-determinism in the system. Processes may vary in speeds, unable to determine the order of originated messages that they received, and due to the arbitrary behavior of faulty processes.

A protocol reaches consensus if it satisfies that every honest process decides on the same value, after a finite number of steps.
There are two theoretical models for the Byzantine consensus problem; they are \emph{Byzantine broadcast} and \emph{Byzantine agreement}~\cite{pease1980reaching,Lamport82}. In Byzantine broadcast, a designated sender tries to broadcast a value to the processes; whereas in Byzantine agreement, every process holds an initial input value. Byzantine agreement requires $f < n/3$ under partial synchrony or asynchrony even with digital signatures but can be achieved with $f < n/2$ under synchrony.

For Byzantine agreement~\cite{Lamport82,rodrigues2001base}, each process initiates a Byzantine broadcast to send its value to peers in parallel. After the broadcast, every honest process will share the same vector of values $V$ = \{$v_1$,$v_2$,$\dots$,$v_h$\}, where $v_i$ is the input value of an honest process $i$. If all honest proceses start with the same input value $v$, then $v$ will be the most frequent in $V$, achieving validity condition. Agreement is reached since all honest processes share the same $V$.

\section{Common Knowledge: Probabilistic Byzantine Fault Tolerance} \label{se:ourapproach}

In this section, we present our approach to study BFT systems in which cheaters have probability of cheating. 

\subsection{Problem definition}

A network is comprised of $n$ processes $p_1, p_2, \dots, p_n$, where $n > 3$. Each process $p_i$ can also be denoted by $i$. The processes communicate with each other via messages. 
We consider a BFT system with $f$ Bazyntine processes and 1-$f$ Zenta processes. The upper bound of $f$ may vary between different systems; for example, $\lfloor \frac{n-1}{3} \rfloor$ in some systems and $\lfloor \frac{n-1}{2} \rfloor$ in others.

At each day $d$, there is a question $q$ given to all processes and 
each process $i$ needs to give a binary answer $a_i(q)$. Assume that all processes except Byzantine ones are honest and well-behaved. They can always give correct answer, i.e., $\bf{q}$. However, some process $i$ in $t$ may sometimes give a wrong answer, e.g., $a_i(q)$ = $!{\bf q}$, with a probability of $\epsilon_i$ where $0 \leq \epsilon_i \leq 1$.
For each honest process $i$, its $\epsilon_i$ is 0. Whilst a Byzantine process $i$ has a $\epsilon_i$ greater than zero.

\subsection{Answer vector and common state}
At day $d$, a question $q_d$ is given to the processes. The answers from the processes are given by the vector: 
$a^{(d)} = [a_1(q_d), a_2(q_d), \dots, a_n(q_d)]$.
The common answer, which supermajority agree on, is guaranteed the correct answer by BFT systems. Any voter who gives an answer that is different from the common answer is a cheater.

Let $\overline{a^{(d)}}$ be the mean value of the vector $a^{(d)}$.
There are two cases in 1/3-BFT systems: (a) if supermajority of the answers is 1, then $\overline{a} > 2/3$; (b) if supermajority of the answers is 0, then $\overline{a} < 1/3$.
Hence, a process $i$ is a cheater if the difference $a_i(q_d) - \overline{a} > 2/3$.
Thus, every process can deduce from the common answer, which is computed from the vector $a^{(d)}$, to see whether other processes including him/herself are cheaters.

\subsection{Know relation}

Let $c_i(d)$ denote that $i$ is cheating at day $d$. For simplicity, in this section, we consider the cheater $i$ has the same probability of cheating $\epsilon_i$ the same every day. That is
The $c_i(d)$ equals to $\epsilon_i$.

Let $k(i,j)$ denote the certainty level that process $i$ knows that process $j$ be a Bazyntine process.
We make no assumption about whether a process knows itself is Bazyntine or not.

In the first day, the probability that process $i$ knows process $j$ is a cheater is $k(i,j) = 1 - (1 - \epsilon_j) = \epsilon_j$.
On the $d$-th day, process $i$ knows that process $j$ is a cheater or not with a certainty that equals to:
\begin{equation}
k(i,j) = 1 - (1 - \epsilon_j)^d
\end{equation}

As $d$ increases, $(1-\epsilon_i)^d$ reduces toward zero because $\epsilon_j \in [0,1]$. Thus, $k(i,j)$ tends to 1 after sufficiently large $d$ days.

\subsection{Cheating Detection}
We now present a matrix to capture the know relation for every pair of processes. 

Let matrix $K$denote who-knows-whom matrix, in which each value at row $i$ and column $j$ is the value of $k(i,j)$. The value at $K(i,j)$ is equal to $k(i,j)$, which is the certainty level that process $i$ knows that process $j$ is a Bazyntine.

\begin{equation*}
K_{n,n} = 
\begin{bmatrix}
k_{1,1} & k_{1,2} & \cdots & k_{1,n-1} & k_{1,n} \\
k_{2,1} & k_{2,2} & \cdots & k_{2,n-1} & k_{2,n} \\
\vdots & \vdots & \ddots & \vdots & \vdots \\
k_{n-1,1} & k_{n-1,2} & \cdots & k_{n-1,n-1} & k_{n-1,n} \\
k_{n,1} & k_{n,2} & \cdots & k_{n,n-1} & k_{n,n} 
\end{bmatrix}
\end{equation*}

On day one, the matrix $K^{(1)}$ is given by:
 \begin{equation*}
K^{(1)}_{n,n} = 
\begin{bmatrix}
\epsilon_1 & \epsilon_2 & \cdots & \epsilon_{n-1} & \epsilon_n \\
\epsilon_1 & \epsilon_2 & \cdots & \epsilon_{n-1} & \epsilon_n \\
\vdots & \vdots & \ddots & \vdots & \vdots \\
\epsilon_1 & \epsilon_2 & \cdots & \epsilon_{n-1} & \epsilon_n \\
\epsilon_1 & \epsilon_2 & \cdots & \epsilon_{n-1} & \epsilon_n 
\end{bmatrix}
= \epsilon_1 . I_1 + \epsilon_2 . I_2 + \dots + \epsilon_n . I_n = \sum_{i=1}^n{\epsilon_i . I_i},
\end{equation*}

where $I_i$ is the $n \times n$ matrix, whose $i$-th column contains all 1's.
 \begin{equation*}
I_{i} = 
\begin{bmatrix}
	0 & 1 & \cdots & 0 & 0 \\
	0 & 1 & \cdots & 0 & 0 \\
	\vdots & \vdots & \ddots & \vdots & \vdots \\
	0 & 1 & \cdots & 0 & 0 \\
	0 & 1 & \cdots & 0 & 0 
\end{bmatrix}
\end{equation*}

On the $d$-th day, the matrix becomes:
\begin{equation*}
K^{(d)} = 
\begin{bmatrix}
1-(1-\epsilon_1)^d & 1-(1-\epsilon_2)^d & \cdots & 1-(1-\epsilon_{n-1})^d & 1-(1-\epsilon_n)^d \\
1-(1-\epsilon_1)^d & 1-(1-\epsilon_2)^d & \cdots & 1-(1-\epsilon_{n-1})^d & 1-(1-\epsilon_n)^d \\
\vdots & \vdots & \ddots & \vdots & \vdots \\
1-(1-\epsilon_1)^d & 1-(1-\epsilon_2)^d & \cdots & 1-(1-\epsilon_{n-1})^d & 1-(1-\epsilon_n)^d \\
1-(1-\epsilon_1)^d & 1-(1-\epsilon_2)^d & \cdots & 1-(1-\epsilon_{n-1})^d & 1-(1-\epsilon_n)^d 
\end{bmatrix}
= \sum_{i=1}^n{(1-(1-\epsilon_i)^d) . I_i}
\end{equation*}

Since $0 < \epsilon_i \leq 1$, $(1-\epsilon_i)^d$ tends to 0 as $d$ increases. Thus, $1-(1-\epsilon_i)^d)$ tends to 1 for every cheater $i$. That is, for every cheater $i$, the column $i$ of $K^{(d)}$ has all 1's. Remarkably, it shows a common knowledge that every cheater $i$ is known by all processes for some sufficiently large value of $d$ (days).

\section{Extended study: Probabilistic Byzantine Fault Tolerance in aBFT systems} \label{se:asyn}

In this section, we extend our study of the cheater problem in asynchronous BFT systems. We first show a motivation example of the asynchronous cheater problem that captures a simple version of Byzantine fault detection problem in asynchronous system.
We then show our formulation to study common knowledge of Byzantine processes who probabilitically cheats.

\subsection{The asynchronous cheater problem} \label{sec:asyncheater}
Assume there is an village of $n$ villagers in a remote island.
On the first day, a group of $k$ villagers come out to collect fruits and brings them back in a box to the village. Each villager counts the number of fruits they collected and individually records their count on the box. 
On the 2nd day, another group of $k$ villagers will do fruit picking and bringing the fruits back, and they invididually writes down their count. Each of them then finds the boxes collected in previous days and writes his/her count, if it was not done previously. And so on. 
Each villager once writes down their count, they can check if there is a count that majority of them agrees on, and hence s/he can find out whether any of the $k$ is honest or cheating. If a cheater is found, that villager can gossip the new finding with all villagers on that day

Remarkably, the asynchronous version is different from the previous version of the cheater problem (described in~\ref{sec:cheater}). The difference is that on every day, the box of that day only has $k$ ($<n$) counts, whereas previous version all $n$ counts are done on the same day.

\subsection{Problem definition}
Given a network of $n$ processes with $t$ honest processes and $n-t$ Byzantyne processes. At each day, a group of $k$ processes are selected; and each of them is given a question $q$ of the day and is required to give a binary answer $a_i(q)$. They also give their answer to the questions of previous days that they havenot answered yet. 
Each process $i$ may give an wrong answer, e.g., $a_i(q)$ = $\bar{\bf q}$, with a probability of $\epsilon_i$ where $0 \leq \epsilon_i \leq 1$. Honest processes always give correct answer, say $a_i(q)$ = $\bf q$.

For the sake of simplicity of formulation, we assume that each processes are labeled with an index. Processes of the selected group will take turn (based on their index) to write their answer to the question(s).
 
We define a notion of \emph{round} to indicate when a question from the sequence has been answered by all the processes.
With this definition, the formulation can be expressed similarly to the previous version, except we replace day by round.

At round $d$, a question $q_d$ has the answers from all the processes. The answers are given by the vector: 
$$a^{(d)} = [a_1(q_d), a_2(q_d), \dots, a_n(q_d)]$$
Each process can deduce from the vector $a^{(d)}$ the common answer agreed by the supermajority. Each process $i$ can then figure out whether there are cheaters including him/herself. 

\subsection{Know relation}

Let $c_i(d)$ denote that $i$ is cheating at round $d$. We assume that the a cheater $i$ is cheating with the same probability on every day and every round, that is $c_i(d) = \epsilon_i$.
Let $k(i,j)$ denote the certainty level that process $i$ knows that process $j$ be a Bazyntine process. Presumably, a process may not know whether itself is Bazyntine or not.

In the first round, the probability that process $i$ knows process $j$ is a cheater is $k(i,j) = 1 - (1 - \epsilon_j) = \epsilon_j$.
At round $d$, process $i$ knows that process $j$ is a cheater equals to $$k(i,j) = 1 - (1 - \epsilon_j)^d$$

As $d$ increases, $(1-\epsilon_i)^d$ reduces toward zero, and so 
$k(i,j)$ tends to be 1.

\subsection{Cheating Detection}
Similar to the previous version of synchronous cheater problem, eventually all cheaters are found out.

At round $d$, the matrix becomes:
\begin{equation*}
K^{(d)} = 
\begin{bmatrix}
1-(1-\epsilon_1)^d & 1-(1-\epsilon_2)^d & \cdots & 1-(1-\epsilon_{n-1})^d & 1-(1-\epsilon_n)^d \\
1-(1-\epsilon_1)^d & 1-(1-\epsilon_2)^d & \cdots & 1-(1-\epsilon_{n-1})^d & 1-(1-\epsilon_n)^d \\
\vdots & \vdots & \ddots & \vdots & \vdots \\
1-(1-\epsilon_1)^d & 1-(1-\epsilon_2)^d & \cdots & 1-(1-\epsilon_{n-1})^d & 1-(1-\epsilon_n)^d \\
1-(1-\epsilon_1)^d & 1-(1-\epsilon_2)^d & \cdots & 1-(1-\epsilon_{n-1})^d & 1-(1-\epsilon_n)^d 
\end{bmatrix}
= \sum_{i=1}^n{(1-(1-\epsilon_i)^d) . I_i}
\end{equation*}

As $d$ increases, $1-(1-\epsilon_i)^d)$ tends to 0 for every cheater $i$. Thus, the column $i$ of $K^{(d)}$ has all 1's.

\section{Discussions} \label{se:discuss}
In this section, we give a discussion about our approach with respect to the chance of detection of those Byzantine processes. We also discuss about how we can model in case the Byzantine processes vary in their probability every day (round).

\subsection{Probability of Detection}

Let $i$ and $j$ denote two Byzantine processes. Let $\epsilon_i$ and $\epsilon_j$ be their probability of being Byzantine, respectively.
The chance of being discovered for each of them by a process $m$ at day (round) $d$ are given by:
$k(m,i) = 1-(1-\epsilon_i)^d$, and $k(m,j) = 1-(1-\epsilon_i)^d$, respectively.
Hence, the difference of their chances is given by:
\begin{equation}
\begin{aligned}
k(m,i) - k(m,j) 
&= [1-(1-\epsilon_i)^d] - [1-(1-\epsilon_j)^d] \\
&= (1-\epsilon_j)^d - (1-\epsilon_i)^d \\
&= (\epsilon_i-\epsilon_j) \sum_{k=0}^{n}{(1-\epsilon_i)^k . (1-\epsilon_j)^{d-1-k} }
\end{aligned}
\end{equation}

Since $(1-\epsilon_i)^k . (1-\epsilon_j)^{d-1-k} > 0$, the sign of $k(m,i) - k(m,j)$ depends on the sign of $\epsilon_i-\epsilon_j$. Without loss of generality, we assume that $\epsilon_i > \epsilon_j$. This gives $\epsilon_i - \epsilon_j > 0$, and so $k(m,i) - k(m,j) > 0$ from the above equation. Thus, $k(m,i) > k(m,j)$.

Remarkably, we have shown that a Byzantine process with a higher cheating probability has a higher chance of being caught every single day (round) in synchronous (asynchronous) BFT systems.

\subsection{Randomized Byzantine processes}
So far, we have assumed that individual Byzantine process has the same cheating probability $\epsilon_i$ every day (round). We now consider a more general problem in which each Byzantine process has a varying probability of cheating over time.

Let $\epsilon_i^{(l)}$ denote the probability of Byzantine fault of Byzantine process $i$ at day (round) $l$.
The chance of being discovered for process $i$ by a process $m$ at day (round) $d$ is given by: 
\begin{equation}
k(m,i) = 1-\prod_{l=1}^{d}(1-\epsilon_i^{(l)})
\end{equation}

As the number of days (rounds) $d$ increases, $\prod_{l=1}^{d}(1-\epsilon_i^{(l)})$ approaches to 0. Thus, the value of $k(m,i)$ tends to 1.

\subsubsection{Probability of Detection}
For two processes $i$ and $j$, the difference of the chance of being detected is given by:
\begin{equation}
\begin{aligned}
k(m,i) - k(m,j) 
&= [1-\prod_{l=1}^{d}(1-\epsilon_i^{(l)})] - [1-\prod_{l=1}^{d}(1-\epsilon_j^{(l)})] \\
&= \prod_{l=1}^{d}(1-\epsilon_j^{(l)}) - \prod_{l=1}^{d}(1-\epsilon_i^{(l)})
\end{aligned}
\end{equation}

If cheater $i$ is more likely to cheat than cheater $j$, e.g, $\epsilon_i^{(l)} > \epsilon_j^{(l)}$ for all $l$, then we can deduce from the above equation that $k(m,i) - k(m,j) > 0$. Hence, we can come up with a similar conclusion that a randomized Byzantine process with a higher cheating probability has a higher chance of being caught every single day (round) in synchronous (asynchronous) BFT systems.

\section{Conclusion}\label{se:con}
In this paper, we show that the ability to reason about common knowledge of Byzantine processes is crucial in BFT systems. We have presented an approach that uses a semi-formal model to compute the probability of whether the existence of a Byzantine process becomes a common knowledge of all processes. We have addressed the Byzantine fault detection problem using a matrix form of probabilities. This is an important problem to ensure the substainability of trustless systems.

We have found several interesting properties of common knowledge of probabilistic Byzantine processes from the study of our model. As time goes by (either number of days in synchronous case, or number of rounds in asynchronous case), all cheaters in the network are not only be detected, but also their existence is proven a common knowledge for all processes. Intuitively, we have also shown that the higher the probability of cheating by a process, the higher chance that process being found by the network.

\section{Reference}\label{se:ref}

\renewcommand\refname{\vskip -1cm}
\bibliographystyle{abbrv}
\bibliography{bft}

\end{document}